\documentclass[11pt]{article}

\usepackage{graphicx,apalike}%

\newcommand{\etal}{\textit{et al.}}

\newcommand{\ie}{i.e.}

\newcommand{\given}{\! \mid \!}
\newcommand{\phihat}{\hat{\phi}}

\newcommand{\kappahat}{\hat{\kappa}}

\newcommand{\bbn}{Bayesian belief network}
\newcommand{\bbns}{Bayesian belief networks}

\newcommand{\eqref}{\ref}

\newcommand{\doubleint}{\int\!\!\!\int}
\newcommand\tripleint{{\int\!\!\!\int\!\!\!\int}}

\newcommand{\pa}[1]{\ensuremath{\mathrm{Pa}(#1)}} 
\newcommand{\ch}[1]{\ensuremath{\mathrm{Ch}(#1)}}
\newcommand{\set}[1]{\ensuremath{ \left \{#1 \right \}}}

\newcommand{\relstr}{K}

\begin{document}

\title{{\LARGE \bf Neural Propagation of Beliefs}}
\author{\bf 
    M.J.~Barber\thanks{Institut f\"ur Theoretische Physik, 
    Universit\"at zu K\"oln,
    D-50937 K\"oln,
    Germany} \and 
    \bf J.W.~Clark\thanks{ Department of 
    Physics, Washington University, Saint Louis, MO 63130}    \and
    \bf C.H.~Anderson\thanks{Department of Anatomy and Neurobiology, Washington University 
School of Medicine,  Saint Louis, MO 63110}
}

\date{\today}    
\maketitle

\abstract{We continue to explore the hypothesis that neuronal  
populations represent and process analog variables
in terms
of probability density functions (PDFs).  
A 
neural assembly encoding the joint
probability density over  relevant analog variables can in principle 
answer any meaningful 
question about these variables by implementing the
Bayesian rules of inference.  
Aided by an intermediate representation of the probability 
density based on orthogonal functions spanning an underlying 
low-dimensional function space, 
we show how neural circuits may be 
generated from Bayesian belief networks.  The ideas and the formalism 
of this PDF approach are illustrated and tested with several 
elementary examples, and in particular
through a problem in which model-driven
top-down information flow influences the processing of bottom-up 
sensory input.
}

\section{Introduction}

\subsection{Fundamental Hypothesis}

This is the second of two papers elaborating upon the proposition
\cite{anderson:1994} that 
neural populations encode and process information about analog 
variables in the form of probability density functions (PDFs).  
As demonstrated in 
the first paper \cite{barber/clark/anderson:2001a}, explicit 
representation of probabilistic descriptors of the state of knowledge 
of physically relevant variables subserves a powerful strategy for 
modeling neural circuits.  By exploiting mathematical 
tools developed within 
 the theory of Bayesian inference, we can 
establish general procedures for building and understanding
 models of cortical circuits that carry 
out well-posed information-processing tasks.

Importantly, if 
neural populations encode the joint probability distribution over the 
variables of interest, then such neural networks are able to answer 
any probabilistic question about those variables.  Motivated by this 
fact, we shall now extend
the PDF
hypothesis by devising 
methods for embedding joint probabilities into neural networks.  This 
will enable us to construct neural circuit models that pool multiple sources 
of evidence, such as sensory inputs and any evolutionarily determined 
priors on the joint distribution.  More specifically, we will focus on 
developing neural networks that use ``bottom-up'' sensory inputs to 
build an internal model of the data, which in turn uses ``top-down'' 
signals to impose global regularities on the sensory data.  In the 
resulting neural networks, there will naturally arise
distinct feedforward, feedback, and lateral connections, analogous to the neural 
pathways observed in the anatomy of the cerebral cortex 
\cite{vanessen/etal:1992}.  


Our treatment is based on  Bayesian belief 
networks \cite{pearl:1988,smyth/etal:1997}, 
graphical representations of 
probabilistic models that provide an efficient means for organizing the
relations between the random variables of a given model.  The resulting neural networks 
will have several properties of \bbns, as well as more typical neural-network 
properties, so we will call them \emph{neural belief 
networks}.  Bayesian belief network have been previously 
utilized in the genesis of certain  neural networks 
(\cite{neal:1992,zemel:1999}), although with different methods for 
generating the neural network architecture and dynamics.

\subsection{Three Levels of Representation}\label{sec:repxforms}

In the context of the PDF hypothesis \cite{anderson:1996}, we  assert 
that a physical variable $x$ is described 
by a neural population at time $t$ in terms of a PDF 
$\rho(x;t)$, rather than as a single-valued estimate $x(t)$.  In 
general, we consider a PDF described at time $t$ in terms of a 
set of $D$ parameters \set{A_{\mu}}. 
  Experimentally observed linear decoding rules 
\cite{georgopoulos/etal:1986,schwartz:1993} suggest that the PDFs may 
be represented as 
\begin{equation}
    \rho(x;t) = \sum_{\mu=1}^{D}A_{\mu}(t) \Phi_{\mu}(x)
    \label{eq:definerep}
\end{equation}
The basis functions $\Phi_{\mu}(x)$ are orthonormal functions which 
serve to define 
the PDFs that the neural circuit can represent, but
the amplitudes $A_{\mu}(t)$ 
cannot be interpreted as neuronal firing 
rates due to their arbitrarily high precision and their ability to take 
on negative values.  Therefore, we introduce an additional  
representation of the PDF in terms of firing rates $a_{i}(t)$ and
decoding functions $\phi_{i}(x)$ 
assigned to $N$ neurons, so that
\begin{equation}
    \rho(x;t) = \sum_{i=1}^{N}a_{i}(t)\phi_{i}(x)
    \label{eq:defineneurrep}
\end{equation}
Unlike the basis functions \( \Phi_{\mu}(x) \), the decoding functions 
\( \phi_{i}(x) \)
form a highly redundant, overcomplete representation 
(\( N \gg D \)) suitable for use with neuronal units having 
biologically realistic precision (some 2--3 bits; see Rieke 
\etal, 1997\nocite{rieke/etal:1997}).  

The abstract representation defined by equation~\ref{eq:definerep} will 
underlie the representation in the neuron space defined by 
equation~\ref{eq:defineneurrep}.  This allows us to deal with the issue of how 
PDFs can be precisely implemented in populations of neurons by focusing 
on the mapping between the minimal space and the space of neurons 
\cite{barber/clark/anderson:2001a}.  
Thus, neural belief networks can be developed in the 
theoretically convenient abstract representation, and then be implemented 
in more realistic networks of low-precision model neurons.  
Adopting the terminology of  Zemel \etal\ (1998\nocite{zemel/etal:1998}), we  denote the 
set of physical 
variables as the \emph{implicit space} and the measurable quantities 
as the \emph{explicit space}.  Extending their nomenclature, we  
denote the abstract space of equation~\ref{eq:definerep} as the 
\emph{minimal space}.  
The explicit space of neurons constitutes a biological implementation of the 
desired computations in the implicit space, 
while the minimal space, whose properties are more conducive to 
formal analysis, provides a valuable
bridge between the two other spaces.


We have developed rules to 
transform between representations in the three spaces 
\cite{barber/clark/anderson:2001a}.  
Equations~\eqref{eq:definerep}
and~\eqref{eq:defineneurrep} respectively define how to transform 
representations in 
the minimal and explicit spaces into a representation (PDF) in the 
implicit space.  With the remaining rules (summarized below), 
we can readily switch 
between the three representations and approach any task in the 
most appropriate space.

For the orthonormal basis \( \set{\Phi_{\mu}(x)} \), the 
minimal space coefficients \( A_{\mu}(t) \) are found using the 
encoding rule
\begin{equation}
    A_{\mu}(t) = \int \Phi_{\mu}(x) \rho(x;t) \, dx
    \label{eq:iuxform}
\end{equation}
The coefficients in the explicit space, \ie\ the neural firing rates \( 
a_{i}(t) \), cannot be found in this direct fashion.  We utilize a 
set of encoding functions \( \phihat_{i}(x) \) and an encoding rule 
of the form
\begin{equation}
    a_i(t) = f \left (\int \phihat_i(x) \rho(x;t)\, dx \right )
    \label{eq:iexform}
\end{equation}
This explicit-space encoding rule is patterned after the neural 
responses associated with the population vector of Georgopoulos 
\etal\ (1986)\nocite{georgopoulos/etal:1986}.

To relate the explicit and minimal spaces, we  express the 
encoding and decoding functions in terms of the orthonormal basis, so 
that 
\begin{eqnarray}
    \phi_{i}(x) = \sum_{\nu=1}^{D} \kappa_{\nu 
    i}\Phi_{\nu}(x)\label{eq:Phitophi}\\
    \phihat_{i}(x) = \sum_{\nu=1}^{D}\kappahat_{i\nu}\Phi_{\nu}(x)
\end{eqnarray}
The transformation coefficients \( \kappa_{\nu i} \) and \( \kappahat_{i\nu} \) also 
relate the \( A_{\mu}(t) \) and \( a_{i}(t) \), such that
\begin{eqnarray}
    A_{\mu}(t) &=& \sum_{i=1}^{N}\kappa_{\nu i}a_{i}(t) 
    \label{eq:euxform}\\
    a_{i}(t) &=& f \left ( \sum_{\nu=1}^{D}\kappahat_{i\nu} A_{\nu}(t) \right )
    \label{eq:uexform}
\end{eqnarray} 
Methods for  determining the encoding functions \( \phihat_{i(x)} 
\), decoding functions \( \phi_{i}(x) \), and transformation coefficients 
\( \kappa_{\nu i} \) and \( \kappahat_{i\nu} \) have been addressed 
in detail in the earlier work 
\cite{barber/clark/anderson:2001a}.

\section{Neural Belief Networks}\label{sec:bbnchap}

\subsection{Bayesian Belief Networks}

In section~\ref{sec:repxforms}, we have summarized methods for encoding and 
decoding probability 
density functions into and from the firing rates of populations of neurons.  
These methods are oriented towards encoding  a single random variable 
(or vector), 
but we do not wish to restrict ourselves to 
only the simplest implicit spaces.
In this work, we will explore ways in which we can 
apply the methods so far developed \cite{barber/clark/anderson:2001a} 
to more complicated implicit spaces.  In 
particular, we will use Bayesian belief networks 
to efficiently organize the implicit random variables, and then 
use these \bbns\ to generate neural networks.

\begin{figure}[tbp]
    \centering
    \includegraphics[width=3in]{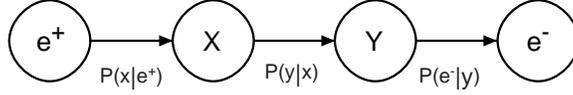}
    \caption{A chain-structured Bayesian belief network.  Evidence $e^{+}$ 
    and $e^{-}$ from the two ends of the chain influences the 
    belief in the random variables $X$ and $Y$.  In 
    a straightforward
    terminology, $X$ is referred to as the parent of $Y$ and $Y$ as the child 
    of $X$.  From the structure of the graph, we can see for example that $Y$ 
    is conditionally independent of $e^{+}$ given $X$; this is true 
    regardless of the values of the links $P(e^{-}\given y)$, $P(y\given 
    x)$, and $P(x\given e^{+})$.  }
    \label{fig:exyechain}
\end{figure}

Bayesian belief networks are directed acyclic graphs that represent 
probabilistic models (Figure~\ref{fig:exyechain}).  
Each node represents 
a random variable, and the arcs (or directed line segments) 
signify the presence of direct causal 
influences between the linked variables.  The strengths of these 
influences are defined using conditional probabilities.  The 
directionality of a specified link indicates the direction of causality (or, 
more simply, relevance); an arc points from direct cause to effect.

\bbns\ have two properties that we will find very useful, both of which stem 
from the independencies shown by the graph structure.  First, the value 
of a node $X$ is not dependent upon all of the other graph nodes.  
Rather, it depends only on a subset of the nodes, called a Markov 
blanket of $X$, that separates node $X$ from all the other nodes in 
the graph.  The Markov blanket of interest to us can be readily determined from 
the graph structure.  It is comprised of the union of the direct parents 
of $X$, the 
direct successors of $X$, and all direct parents of the direct 
successors of $X$.  Second, the joint probability over the random 
variables is decomposable as
\begin{equation}
    P(x_{1},x_{2},\ldots,x_{n}) = \prod_{i=1}^{n} P(x_{i}\given \pa{x_{i}})
\end{equation}
where \pa{x_{i}} refers to the 
(possibly empty)
set of direct parent nodes of $X_{i}$.  
This decomposition comes about from repeated application of Bayes' 
rule and from the structure of the graph.  

\subsection{Probabilistic Inference Performed by Neural Networks} \label{sec:inferencenbns}

Before exploring arbitrary Bayesian belief networks, 
it is enlightening to consider a BBN with a simple graph 
consisting of two 
connected nodes \( X \longrightarrow Y \).
This  graph 
represents any probabilistic model where a single random variable is 
inferred from one source of evidence.
For convenience, we will work in the minimal space.  

Our objective 
is to find the
best
marginal PDFs $\rho(y;t)$ and $\rho(x;t)$ 
to describe the system.  We represent the PDFs using 
equation~\ref{eq:definerep} and 
\begin{equation}
    \rho(y;t)  =  \sum_{\nu}B_{\nu}(t) \Psi_{\nu}(y)
    \label{eq:xyinfrepy}
\end{equation}
In this network, we find values of the output firing rates 
\set{B_{\nu}(t)} only, with input firing rates \set{A_{\mu}(t)}
assumed to be fully 
determined by an encoding process.

We define a cost function by
\begin{eqnarray}
    E_{y} &=& \frac{1}{2}\int \left ( \rho(y;t) - 
    \int \rho(y\given x) \rho(x;t) dx 
    \right )^{2}dy \nonumber\\*
    &=& \frac{1}{2}\int \left ( \sum_{\nu}B_{\nu}(t) \Psi_{\nu}(y) -  
    \sum_{\mu}A_{\mu}(t) \int \rho(y\given x)\Phi_{\mu}(x) dx \right 
    )^{2}dy \nonumber\\*
    && \label{eq:xyinferencecostfunction}
\end{eqnarray}
The cost function has a minimum corresponding to  
 taking a weighted 
average of the conditional probability $\rho(y\given x)$
\begin{equation}
\rho(y;t) = \int \rho(y\given x) \rho(x;t) dx
\label{eq:justifyinference}
\end{equation}
We  assume in equation~\ref{eq:justifyinference} that the 
relationship between $x$ and $y$ is independent of the values of the 
underlying parameters \( A_{\mu}(t) \) of the minimal space
\cite{barber/clark/anderson:2001a}.

To minimize \( E_{y} \), we 
calculate the derivatives \( {\partial E_{y}} / {\partial B_{\nu}} \).
Using gradient
descent,
we obtain the update rule
\begin{eqnarray}
     \frac{dB_{\nu}(t)}{dt} & = & -\eta\frac{\partial E_{y}} 
    {\partial B_{\nu}}\nonumber\\   
    & = & -\eta\left( B_{\nu}(t) - \sum_{\mu} 
    \Omega_{\nu\mu}A_{\mu}(t) \right)
    \label{eq:bbnxyinf}
\end{eqnarray}
where \( \eta \) is a rate constant and we identify weights
\begin{equation}
    \Omega_{\nu\mu} = \doubleint \Psi_{\nu}(y) 
    \rho(y\given x)\Phi_{\mu}(x) dx dy \label{eq:bbnxyinfweights}
\end{equation}
Equations~\ref{eq:bbnxyinf} and~\ref{eq:bbnxyinfweights} define a 
network in the minimal space which can be converted to a neural 
network in the explicit space using equation~\ref{eq:uexform}.

Equation~\ref{eq:bbnxyinf} has a fixed point at
\begin{equation}
    B_{\nu}(t) = \sum_{\mu}A_{\mu}(t)\doubleint \Psi_{\nu}(y) \rho(y\given 
    x) \Phi_{\mu}(x) dx dy
    \label{eq:bbnxyinffixedpts}
\end{equation}
This is identical to the result obtained by encoding the inference 
relation in equation~\ref{eq:justifyinference} into the minimal 
space using equation~\ref{eq:iuxform}.

\subsection{Predictive and Retrospective Support}\label{sec:retrosupport}

Neural belief networks can feature two distinct types of information 
propagation that provide support for the PDFs represented at each 
graph node.  Predictive support, also called causal support, is probabilistic 
information that propagates, along the directions of the graph links, from 
cause to effect \cite{pearl:1988}.  The  network considered in 
section~\ref{sec:inferencenbns} involves only predictive support.

The second type of support is retrospective, or diagnostic, support.  
In this case, information propagates against the directions of the graph 
links, from effect to cause, or, equivalently, from evidence to 
hypothesis \cite{pearl:1988}.  
By specifying $\rho(y;t)$ instead of $\rho(x;t)$, the $X 
\longrightarrow Y$ inference network features only retrospective 
support.

Using the same minimal-space representation (equations~
\ref{eq:definerep}
and~\ref{eq:xyinfrepy})  that we used 
for the predictive network, we determine the update rule for the 
retrospective network.  We find
\begin{equation}
    \frac{dA_{\mu}(t)}{dt} = \eta \left ( \sum_{\beta} B_{\beta}(t) 
    \Omega_{\beta\mu} - \sum_{\alpha} A_{\alpha}(t) 
    \Upsilon_{\alpha\mu}\right )
    \label{eq:xyinfretronet}
\end{equation}
with feedback weights
\begin{equation}
    \Omega_{\beta\mu} = \doubleint \Psi_{\beta}(y) \rho(y\given x) 
    \Phi_{\mu}(x) dx dy
    \label{eq:xyinfretrofbwts}
\end{equation}
and lateral weights
\begin{equation}
    \Upsilon_{\alpha\mu} = \int \left(\int\rho(y \given x) 
    \Phi_{\alpha}(x) dx\right) \left(\int\rho(y \given x) 
    \Phi_{\mu}(x) dx\right) dy
    \label{eq:xyinfretrolatwts}
\end{equation}
While the feedback weights~\( \Omega_{\beta\mu} \) of the 
retrospective network
are identical to the feedforward 
weights of the predictive network (equation~\ref{eq:bbnxyinf}) 
in the minimal space, note 
that the 
resulting neural networks weights in the explicit space need not be identical.
The lateral weights~\( \Upsilon_{\alpha\mu} \) provide a measure of 
the correlation between what the 
different basis functions in the parent node $X$ predict about the 
child node $Y$; these lateral connections act to ensure consistency 
between evidence and hypothesis.

\begin{figure}[tbp]
    \centering
    \begin{tabular}{l}
	a  \\
	\includegraphics[width=2.25in]{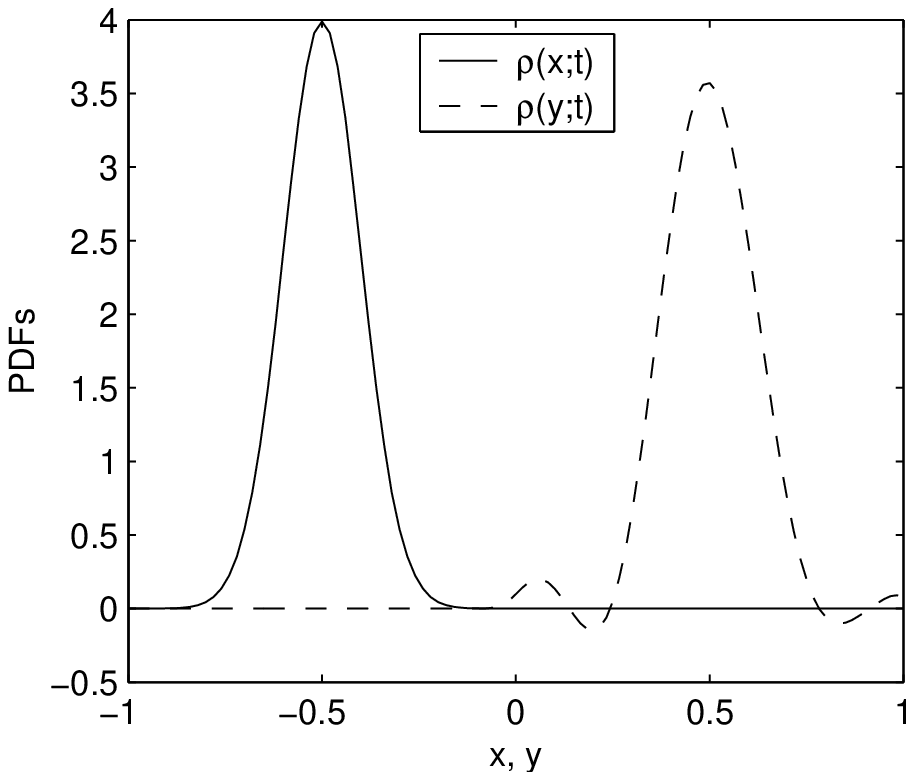}  \\
    \end{tabular}
    \hfill
    \begin{tabular}{l}
	b  \\
	\includegraphics[width=2.25in]{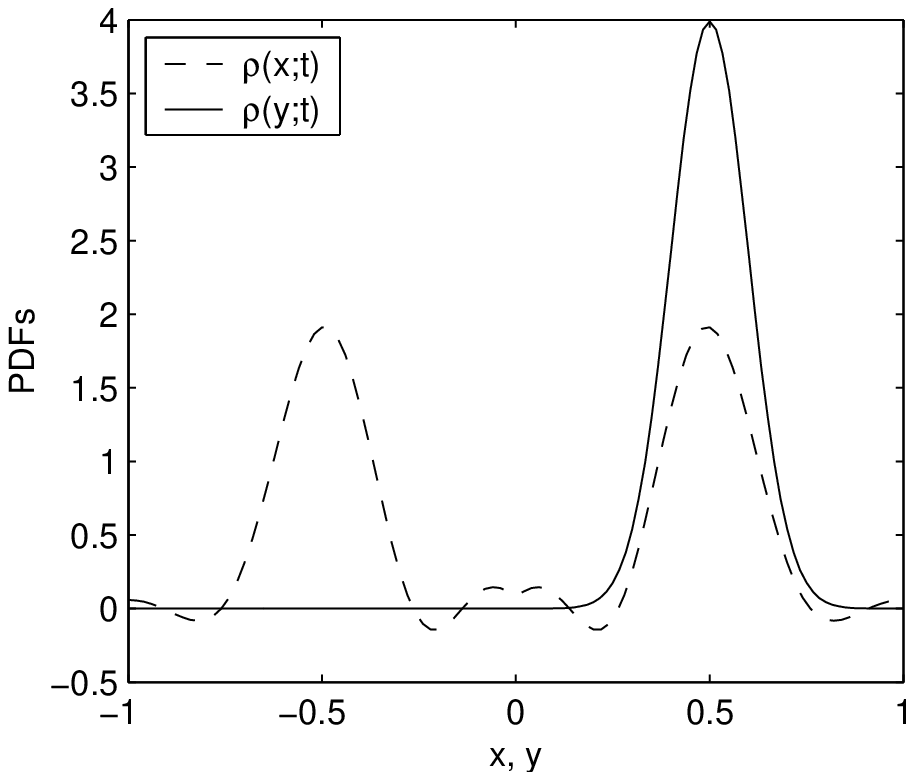}  \\
    \end{tabular}
    \caption{Predictive and retrospective support of the absolute 
    value function.  The results shown here utilize spaces of dimension $ 
    D=6 $ to represent both $x$ and $y$ in the two networks.  
    The structure of the spaces was determined from the singular value 
    decomposition 
    \protect\cite{goldberg:1991}
    of  a discrete approximation 
    of $\rho(y\given x)$---the basis functions $\Phi_{\mu}(x)$ 
    and~$\Psi_{\nu}(y)$ are set to the singular vectors 
    corresponding to the $D$ largest singular values.  (a) 
    The network with predictive support closely approximates the 
    absolute value function.  Specifying a PDF~$\rho(x;t)$ centered 
    about $x=-1/2$  yields an inferred PDF~$\rho(y;t)$ of similar form 
    centered about $x=+1/2$.  (b)  The network with retrospective 
    support allows for both possible solutions.  Specifying a 
    unimodal PDF~$\rho(y;t)$ centered about $x=+1/2$ 
    yields an inferred bimodal PDF~$\rho(x;t)$, 
    with the modes centered about $x=-1/2$ and $x=+1/2$.  }
    \label{fig:xyinfpredretro}
\end{figure}

Although the networks driven by retrospective support are closely 
related to the networks driven by predictive support,  
their function is quite different.
For example, consider a network with 
\( \rho(y \given x) = \delta\left(y - 
\left | x \right |\right) \) as the underlying computation.  In a 
predictive network, wherein we specify \( \rho(x;t) \) and infer \( 
\rho(y;t) \), the absolute-value relationship is approximated in a 
straightforward fashion (Figure~\ref{fig:xyinfpredretro}a).  
Conversely, a retrospective network based upon the same conditional 
PDF is called upon to ``invert'' the absolute value, a non-invertible function.  
The inferred  \( \rho(x;t) \) decoded from the retrospective 
network (Figure~\ref{fig:xyinfpredretro}b) 
captures both of the possible solutions---with positive and 
negative values---in response to a unimodal PDF \( \rho(y;t) \).

\subsection{Encoding Bayesian Belief Networks Into Neural 
Networks}\label{subsec:makinggeneralnbns}

Following a  strategy similar to that presented in section~\ref{sec:inferencenbns}, we can  
develop neural-network update rules from 
arbitrary \bbns.  We assume that the \bbn\ consists of $R$ nodes, 
symbolizing 
random variables $X_{1}$, $X_{2}$, $X_{3}$, \ldots,$X_{R}$.  
Introducing representations 
\begin{equation}
    \rho(x_{i};t) = \sum_{\mu}A^{(i)}_{\mu}(t) \Phi^{(i)}_{\mu}(x_{i})
\end{equation}
for the marginal distributions in the minimal space, we define auxiliary cost 
functions, 
analogous to that in equation~\ref{eq:xyinferencecostfunction}, with 
the forms
\begin{eqnarray}
    E_{i} & = & \frac{1}{2}\int \left(\rho(x_{i};t) - \int\rho(x_{i} \given 
    \pa{x_{i}}) \prod_{j} \rho(x_{j};t) dx_{j} \right)^{2} dx_{i} 
    \nonumber\\
     & = & \frac{1}{2}\int \left({\displaystyle \sum_{\mu}A^{(i)}_{\mu}(t) 
     \Phi^{(i)}_{\mu}(x_{i}) -\mbox{}\hfill \atop \quad 
     \displaystyle \int\rho(x_{i} \given \pa{x_{i}}) {\displaystyle \prod_{j} 
     \sum_{\nu_{j}}}A^{(j)}_{\nu_{j}}(t) \Phi^{(j)}_{\nu_{j}}(x_{j})dx_{j} 
     }\right)^{2} dx_{i}\nonumber\\
     &&
     \label{eq:generalenergies}
\end{eqnarray}
where the index $j$ for the product runs over the direct parents of 
$X_{i}$.  

We further define an aggregate cost function
\begin{equation}
    E =  \sum_{i=1}^{R}\relstr_{i}E_{i}
\end{equation}
The parameters \( \relstr_{i} \) can be used to emphasize 
particular portions of the network; except where otherwise noted, we 
assume that \( \relstr_{i} = 1 \) for all \( i \).
Employing gradient descent to minimize \( E \), we find
\begin{equation}
    \frac{dA_{\sigma}^{(k)}}{dt} = 
    -\eta\sum_{i=1}^{R}\relstr_{i}\frac{\partial E_{i}} {\partial 
     A_{\sigma}^{(k)}}
\end{equation}
Since $A_{\sigma}^{(k)}$ does not appear in all of the cost functions, 
${\partial E_{i}}/{\partial A_{\sigma}^{(k)}}$ is nonzero only for 
$i = k$ and when the graph node for $X_{i}$ is one of the direct 
children of the graph node for $X_{k}$.  Thus,
\begin{equation}
    \frac{1}{\eta} \frac{dA_{\sigma}^{(k)}}{dt} = -\relstr_{k}\frac{\partial 
    E_{k}}{\partial A_{\sigma}^{(k)}} - \sum_{i\in\set{i:X_{i}\in 
    \ch{X_{k}}}} \relstr_{i}\frac{\partial E_{i}}{\partial A_{\sigma}^{(k)}}
    \label{eq:genbbnrulemustevalderivs}
\end{equation}
As was the case for the two-node inference network, the input PDF or 
PDFs will be specified by an encoding process rather than an update 
rule of this sort.  

The derivatives in equation~\ref{eq:genbbnrulemustevalderivs} are 
straightforward but lengthy to evaluate.  We omit the details of the 
calculation and simply state the results
\begin{eqnarray}
    \lefteqn{\frac{1}{\eta} \frac{dA_{\sigma}^{(k)}}{dt} =}\nonumber \\*
    &&-\relstr_{k}\left(A_{\sigma}^{(k)} - \prod_{j} 
    \sum_{\nu_{j}}A_{\nu_{j}}^{(j)} 
    \Omega^{(k)}_{\sigma\nu_{1}\nu_{2}\cdots\nu_{j_{max}}} \right ) 
    \nonumber \\*
     & & \mbox{} + \sum_{i}\relstr_{i}  \sum_{\mu}A_{\mu}^{(i)} \prod_{j \neq 
     k}\sum_{\nu_{j}}A_{\nu_{j}}^{(j)} 
     \Omega^{(ik)}_{\mu\nu_{1}\nu_{2}\cdots\nu_{j_{max}}\sigma} \nonumber \\*
     & & \mbox{} -\sum_{i}\relstr_{i}\prod_{j}\prod_{j' \neq 
     k}\sum_{\nu_{j}}A^{(j)}_{\nu_{j}} \sum_{\mu_{j'}}A^{(j')}_{\mu_{j'}} 
     \Upsilon^{(jj'k)}_{\nu_{1}\nu_{2}\cdots\nu_{j_{max}}\mu_{1}\mu_{2}\cdots 
     \mu_{j'_{max}}\sigma}
    \label{eq:biguglybbn}
\end{eqnarray}
where we have defined
\begin{eqnarray}
    \lefteqn{\Omega^{(k)}_{\sigma\nu_{1}\nu_{2}\cdots\nu_{j_{max}}} = } 
    \nonumber \\
    &&\int \Phi_{\sigma}^{(k)}(x_{k})\rho(x_{k}\given 
    \pa{x_{k}})\prod_{j}\Phi_{\nu_{j}}^{(j)}(x_{j}) dx_{j}dx_{k} 
    \label{eq:biguglywtone} \\
    \lefteqn{\Omega^{(ik)}_{\mu\nu_{1}\nu_{2}\cdots\nu_{j_{max}}\sigma} = 
    }\nonumber \\
    &&\int \Phi_{\mu}^{(i)}(x_{i})\rho(x_{i}\given \pa{x_{i}})  \Phi_{\sigma}^{(k)}(x_{k}) \prod_{j \neq k}\Phi_{\nu_{j}}^{(j)}(x_{j}) 
     dx_{j}dx_{i}dx_{k}
    \label{eq:biguglywttwo} 
\end{eqnarray}
and
\begin{eqnarray} 
    \lefteqn{\Upsilon^{(jj'k)}_{\nu_{1}\nu_{2}\cdots\nu_{j_{max}}\mu_{1}\mu_{2}
    \cdots 
    \mu_{j'_{max}}\sigma} =}\nonumber \\
    &&\int \left ( \int \Phi^{(k)}_{\sigma}(x_{k}) \rho(x_{i}\given\pa{x_{i}}) 
    \prod_{j' \neq k} \Phi^{(j')}_{\mu_{j'}}(x_{j'}) dx_{j'} dx_{k}\right. 
    \nonumber \\
     &&\qquad\qquad\qquad\qquad \left.  \int 
     \rho(x_{i}\given\pa{x_{i}})\prod_{j}\Phi^{(j)}_{\nu_{j}}(x_{j})dx_{j} 
     \right ) dx_{i}
    \label{eq:biguglywtthree}
\end{eqnarray}
The sums over the index $i$ in equation~\ref{eq:biguglybbn} run over 
the children of $X_{k}$.  The products are over the parents of either 
node $X_{k}$ (equation~\ref{eq:biguglywtone}, first term 
in~\ref{eq:biguglybbn}) or node $X_{i}$ (equations~\ref{eq:biguglywttwo} 
and~\ref{eq:biguglywtthree}, second and third term 
in~\ref{eq:biguglybbn}), possibly excluding node~$X_{k}$ itself.  From 
these equations, it can be seen that the PDF represented at a node $X_{i}$ is 
updated based only on its direct parents, its direct descendents, and 
the direct parents of its direct descendents, so that $X_{i}$ is 
separated from all other nodes in the neural belief networks by an 
appropriate Markov blanket.

Equations~\ref{eq:biguglybbn} 
through~\ref{eq:biguglywtthree} can be awkward to use directly.  It 
can be more convenient to write out the cost functions for a specific 
probabilistic model using equation~\ref{eq:generalenergies} and directly 
evaluate the necessary derivatives (specified by 
equation~\ref{eq:genbbnrulemustevalderivs}).

\section{Applications of Neural Belief Networks}

\subsection{Bidirectional Propagation}

So far, we have restricted our attention to developing neural networks 
that encode simple probabilistic models involving only a single source 
of evidence.  Given suitable representations, 
we can design neural networks that 
capture a wide variety of probabilistic relations
\cite{barber/clark/anderson:2001a}.  Further, by 
regarding representations of probabilistic dependence models 
as Bayesian belief networks, we have seen that two types of 
information propagation come into play: predictive and retrospective.

In this section, we will examine several applications of neural belief 
networks.  Unlike those considered before, the probabilistic models will now
feature multiple sources of evidence, with bidirectional propagation of both 
predictive and retrospective support.  The corresponding neural 
networks thus have neurons with both feedforward and feedback 
connections, as well as lateral connections.

\subsection{Neural Propagation of Evidence in Trees}\label{sec:nbntrees}

\begin{figure}[tbp]
    \begin{center}
        \includegraphics[scale=.75]{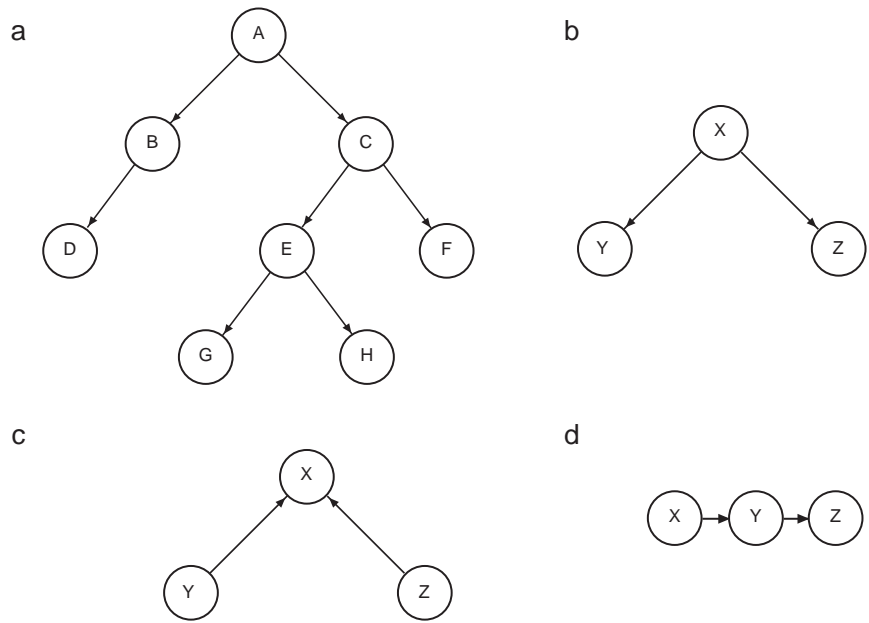}
    \end{center}
    \caption{
    Tree-structured Bayesian belief networks.  
    (a) In this tree, 
    node $A$ is called the root, while nodes $D$, $F$, $G$, and $H$ are 
    called the leaves.  For trees, the direct parent of a node is called its 
    father, and its direct children are called its sons.  Since each 
    father has at most two sons, the tree shown here is a binary tree.
    (b)
    A small tree.  Any of the nodes in this tree can be specified 
    as evidence.  The two leaf nodes could also provide evidence: both leaves 
    can provide information to the root, but if the root and one of the 
    leaves is specified, the other leaf will only be driven by the root 
    (its Markov blanket).  
    (c)
    A tree-like graph with the arrows reversed.  Any two of the 
    nodes can be specified and its information will propagate throughout 
    the network.  Unlike the case of the tree, the Markov blanket 
    of any node is 
    both of the other nodes. 
    (d) Chains are special cases of trees.  This three-node chain can feature both 
    predictive and retrospective support.
    }
    \label{fig:trees}
\end{figure}

To facilitate investigation of information propagation in neural 
belief networks, we first focus on networks whose implicit spaces are 
specified by tree-structured Bayesian belief networks 
(Figure~\ref{fig:trees}a).  
These implicit networks are general enough to 
illustrate the concepts, while yielding neural networks that are readily 
understood.  We will examine binary trees, where each node has at most 
two children, but the results extend simply to more general tree 
structures.

We may assume that evidence is only available in the root node and the 
leaf nodes (although all such nodes need not provide evidence).  
The root node is the single node which has no father and 
is located  at the top of the tree, while the leaf nodes are all the 
nodes which have no children.  
If another
node $X$ were externally specified, the subtree rooted at $X$ could be 
broken off and treated separately.  Conversely, the father node of $X$ 
is unaffected by the descendants of $X$, so they could be deleted from 
the original tree, leaving $X$ as a leaf node.

Clearly, an 
unspecified root node can only receive retrospective support, while 
an unspecified leaf node can only receive predictive support.  All 
other unspecified nodes in the tree will receive both retrospective 
and predictive support.  We will thus need to  consider separately
these three types of nodes when determining the update rules for the 
neural network.  

An unspecified root node $X$ with children $Y$ and $Z$ receives feedback 
inputs (retrospective support) from both of them (Figure~\ref{fig:trees}).
We introduce 
the representations 
\begin{eqnarray}
    \rho(x;t) & = & \sum_{\alpha} A_{\alpha}(t) \Phi_{\alpha}(x) 
    \label{eq:xrepresentation}\\
    \rho(y;t) & = & \sum_{\beta} B_{\beta}(t) \Psi_{\beta}(y) 
    \label{eq:yrepresentation}\\
    \rho(z;t) & = & \sum_{\gamma} C_{\gamma}(t) \Theta_{\gamma}(z)
    \label{eq:zrepresentation}
\end{eqnarray}
Following the procedures described in 
section~\ref{subsec:makinggeneralnbns}, 
the update rule for the root is
\begin{equation}
    \frac{1}{\eta}\frac{dA_{\mu}}{dt} = -\relstr_{y}\frac{\partial E_{y}}{\partial A_{\mu}}
    -\relstr_{z}\frac{\partial E_{z}}{\partial A_{\mu}}
\end{equation}
with
\begin{eqnarray}
    \frac{\partial E_{y}}{\partial A_{\mu}} & = &\sum_{\alpha} A_{\alpha}(t) \int \left(\int\rho(y \given x) 
    \Phi_{\alpha}(x) dx\right) \left(\int\rho(y \given x) 
    \Phi_{\mu}(x) dx\right) dy  \nonumber\\
     & &\mbox{} -\sum_{\beta} B_{\beta}(t) \doubleint \Psi_{\beta}(y) 
     \rho(y\given x) \Phi_{\mu}(x) dx dy
\end{eqnarray}
and
\begin{eqnarray}
    \frac{\partial E_{z}}{\partial A_{\mu}} & = &\sum_{\alpha} 
    A_{\alpha}(t) \int \left(\int\rho(z \given x) \Phi_{\alpha}(x) 
    dx\right) \left(\int\rho(z \given x) \Phi_{\mu}(x) dx\right) dz 
    \nonumber\\
     & &\mbox{} -\sum_{\gamma} C_{\gamma}(t) \doubleint \Theta_{\gamma}(z) 
     \rho(z\given x) \Phi_{\mu}(x) dx dz
\end{eqnarray}
Thus, the firing rates for the root node are driven by a sum of  feedback 
inputs that individually are identical to the input produced by a 
single source of retrospective support 
(section~\ref{sec:retrosupport}).  The
parameters 
$\relstr_{y}$ and $\relstr_{z}$ need not be the same; different values 
may be used to give greater significance to one of the inputs.

The firing rates for an unspecified leaf node are also updated by a 
familiar rule.  Consider a leaf node $X$ with father $U$.  Using 
equation~\ref{eq:xrepresentation} and 
\begin{equation}
    \rho(u;t)  =  \sum_{\delta} D_{\delta}(t) \Lambda_{\delta}(u) 
    \label{eq:urepresentation}
\end{equation}
we obtain the update rule
\begin{equation}
    \frac{1}{\eta} \frac{dA_{\mu}}{dt} = -\relstr_{x} \frac{\partial E_{x}}{\partial 
    A_{\mu}}
\end{equation}
where
\begin{equation}
    \frac{\partial E_{x}}{\partial 
    A_{\mu}} = A_{\mu}(t) - \sum_{\delta}D_{\delta}(t) \doubleint 
    \Lambda_{\delta}(u) \rho(x \given u) \Phi_{\mu}(x) du dx
\end{equation}
This update rule is of course identical to the update rule for the $X 
\longrightarrow Y$ network with predictive support 
(section~\ref{sec:inferencenbns}).

All nonroot, nonleaf nodes have similar update rules.  
For a node $X$ 
with father $U$ and sons $Y$ and $Z$, we impose the representations 
given above in equations~\ref{eq:xrepresentation} 
through~\ref{eq:zrepresentation} and equation~\ref{eq:urepresentation}.
Applying 
the procedure of section~\ref{subsec:makinggeneralnbns} yields an update rule
with form
\begin{equation}
    \frac{1}{\eta} \frac{dA_{\mu}}{dt} = -\relstr_{y}\frac{\partial E_{y}}{\partial A_{\mu}}
    - \relstr_{z}\frac{\partial E_{z}}{\partial A_{\mu}} - \relstr_{x} \frac{\partial E_{x}}{\partial 
    A_{\mu}}
\end{equation}
The partial derivatives of the cost functions are identical to those evaluated 
for the root and leaf nodes.  The descendants and ancestors of $X$ 
thus communicate, in the neural network, only through the intermediary of 
$X$ itself.  This is consistent with the tree structure of the underlying 
Bayesian belief network; given $X$, the descendants of $X$ are conditionally 
independent of the ancestors of $X$ \cite{pearl:1988}.

With these update rules, evidence provided at the root node or at leaf 
nodes will propagate throughout the network.  The manner in which 
evidence is specified will depend on how the probabilistic model is 
posed.  Therefore, the same graph could have different nodes specified for 
different purposes.  For instance, the small tree in 
Figure~\ref{fig:trees}b 
could have any of its nodes represent 
sensory inputs.

If we specify the PDF for the root node, the leaf 
nodes $Y$ and $Z$ will receive predictive support from $X$.  By 
selecting appropriate representations for the PDFs and for the conditional 
probabilities associated with the links, the resulting neural network could subdivide a 
complex, highly general sensory input into simpler, more specialized 
components.  For example, a visual input at a particular retinal 
location might be separated into contrast and color.

Conversely, if we specify PDFs for the leaf nodes, the root node $X$ 
will receive retrospective support from $Y$ and $Z$.  Amongst other 
possibilities, this provides a simple way to model redundant sensory 
inputs.  If we take the conditional probabilities to be of narrow 
Gaussian form, $\rho(y \given x) = N(y; x, \sigma_{y}^{2})$ and 
$\rho(z \given x) = N(z; x, \sigma_{z}^{2})$, then the firing rates 
representing $\rho(x;t)$ will be updated so as to pool the diagnostic 
information from both the sensory inputs, $Y$ and $Z$.

Similar arguments apply to larger trees.  Additionally, larger 
trees may well have the root node and leaf nodes specified simultaneously 
(which is not of interest for the small tree discussed above).  These 
nodes may correspond to sensory inputs, or can represent priors that 
are built into the neural network.  

It is important to recognize that the neural update rules developed 
above apply only to the binary trees.  If the arrows in the binary 
tree graphs are reversed, rather different neural 
update rules are produced.  For example, the directions of the small
tree shown in Figure~\ref{fig:trees}b can be reversed, as shown in
Figure~\ref{fig:trees}c.
Specifying $Y$ and $Z$ yields the multiplicative update rule
\begin{eqnarray}
    \lefteqn{\frac{1}{\eta \relstr_{x}}\frac{dA_{\alpha}(t)}{dt}  = -A_{\alpha}(t) } 
    \nonumber \\ && \mbox{} + \sum_{\beta, \gamma} 
    B_{\beta}(t)C_{\gamma}(t)\tripleint \Phi_{\alpha}(x)\rho(x \given y,z) 
    \Psi_{\beta}(y) \Theta_{\gamma}(z) dx dy dz \nonumber \\
    &&
\end{eqnarray}
while specifying $X$ and $Z$ yields
\begin{eqnarray}
    \lefteqn {\frac{1}{\eta \relstr_{x}}\frac{dB_{\nu}(t)}{dt} = } \nonumber \\*
     & & \quad \sum_{\alpha, \gamma}A_{\alpha}(t)C_{\gamma}(t) \tripleint 
     \Phi_{\alpha}(x)\rho(x \given y, z) \Psi_{\nu}(y) \Theta_{\gamma}(z) 
     dx dy dz \nonumber \\*
    && -  \sum_{\beta, \gamma_{1}, \gamma_{2}} B_{\beta}(t) 
     C_{\gamma_{1}}(t)C_{\gamma_{2}}(t)\int \left \{ \vphantom{\sum_{\alpha, 
    \gamma}}\left ( \doubleint \rho(x \given y, 
     z) \Psi_{\beta}(y)\Theta_{\gamma_{1}}(z) dy dz \right ) 
     \right . \nonumber \\*
     && \left .  \qquad\qquad\qquad\qquad\qquad \left ( \doubleint \rho(x 
     \given y, z) \Psi_{\nu}(y)\Theta_{\gamma_{2}}(z) dy dz \right )  
     \vphantom{\sum_{\alpha, \gamma}} \right \} dx\nonumber\\*
     &&\label{eq:reversetreeretro}
\end{eqnarray}
This latter update rule features a feedforward term that is 
multilinear in the firing rates \set{A_{\alpha}(t)} and \set{C_{\gamma}(t)}.
It also features a nonlinear lateral combination of the firing rates 
\set{B_{\beta}(t)} and \set{C_{\gamma}(t)} which serves to ensure that the 
two parent nodes~$Y$ and~$Z$ are mutually consistent with the PDF of the child 
node~$X$.  If only $X$ is specified, there will be an additional update 
rule for the firing rates \set{C_{\gamma}(t)} that is similar in form to 
equation~\ref{eq:reversetreeretro}.

Although the update rules are more complicated with the directions 
reversed in this manner, the probabilistic model may demand it.  For 
instance, the \bbn\ in Figure~\ref{fig:trees}c is appropriate for 
implementing the arithmetic operations (add, subtract, multiply, and divide).

\begin{figure}[tbp]
    \centering
    \begin{tabular}{l}
	a \\
        \includegraphics[width=2.25in]{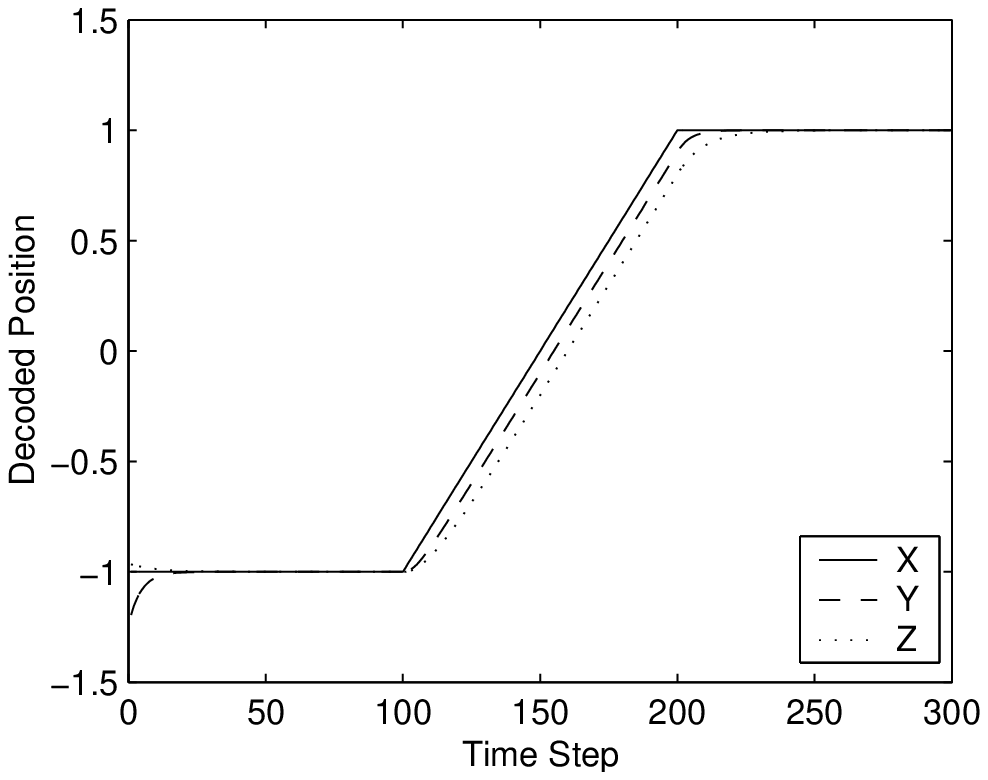}  \\
    \end{tabular}
    \hfill
    \begin{tabular}{l}
	b \\
        \includegraphics[width=2.25in]{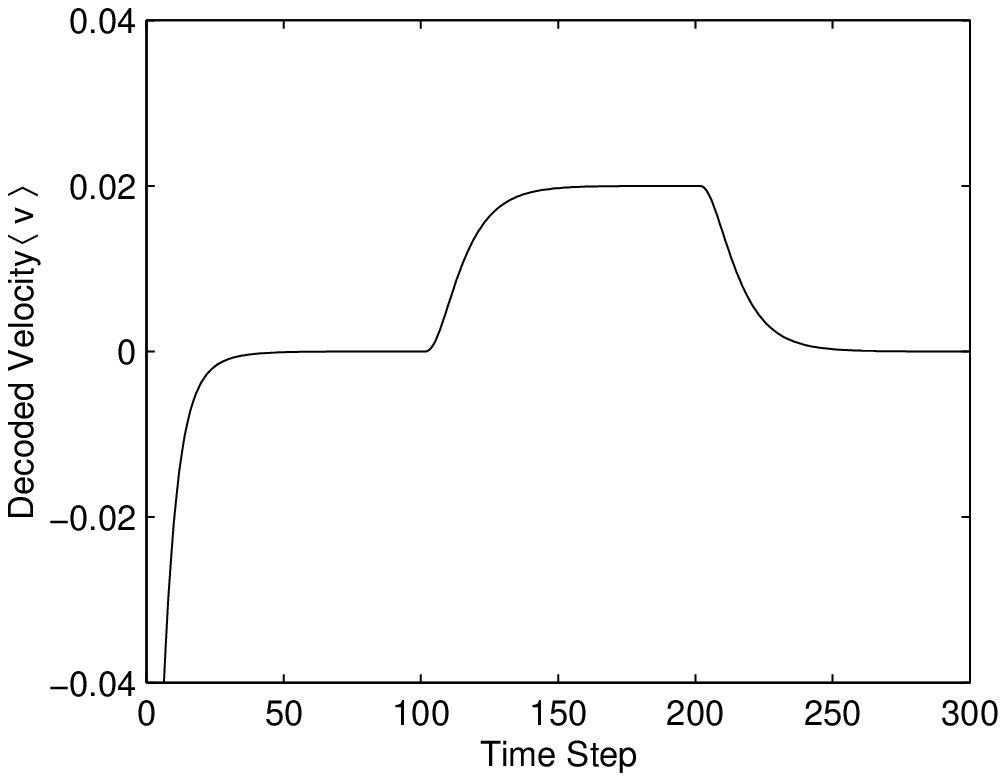}  
    \end{tabular}
    \caption{A neural belief network can estimate the velocity of a moving 
    target.  (a) The position of the target is copied into two different 
    populations of neurons, with different time delays.  (b) The time 
    delay and the difference of the two copies of position are used to 
    estimate the velocity.  The results shown here were obtained with the 
    PDFs for each of the random variables represented using minimal 
    spaces spanned by two straight-line basis functions. }
    \label{fig:estimatev}
\end{figure}

An interesting
application of these two 
types of neural belief networks (tree and reversed-tree) is the 
estimation of the velocity of a moving object.  A small tree 
(Figure~\ref{fig:trees}b) can generate two copies $Y$ and $Z$ of 
the input position $X$ by taking the conditional probabilities to be 
Dirac delta functions $\delta(y - x)$ and $\delta(z - x)$.  
We can set the parameters \( \relstr_{i} \) 
so that the values of the copies will be held 
for different lengths of time.  
In 
particular, we can establish the relations
\begin{eqnarray}
    \rho(y; t) & = & \rho(x; t - \tau) \\
    \rho(z; t) & = & \rho(x; t - 2 \tau)
\end{eqnarray}
These copies of the position can then be used as the inputs to a \bbn\ 
of the type shown in Figure~\ref{fig:trees}c.
By setting $\rho(x \given y, z) = \delta( x - (y - z) / \tau)$, 
the velocity (given here by \( x \)) at time $t - \tau$ 
can be estimated (Figure~\ref{fig:estimatev}).

\subsection{Top-Down Feedback From a High-Level Model}

In section~\ref{sec:nbntrees}, we examined the neural update rules for 
\bbns\ having binary-tree structure.  By doing so, we examined the 
means by which information propagates throughout the network from one 
or more sources.  In particular, we saw that conditional independence 
in the \bbn\ was preserved in the neural network connectivity.  

We now turn 
from the general rules by which probabilistic information propagates 
in the neural network and investigate in more detail the effects that multiple 
sources of evidence have on the encoded PDFs.  We consider PDFs 
encoded by chains of nodes with evidence provided at one or both ends. 
(Specifying the PDF for any other node breaks the chain into two 
chains that can be treated independently.)  A chain is a special case of a 
tree, so the neural update rules can be obtained as in 
section~\ref{sec:nbntrees}.
\begin{figure}[tbp]
    \begin{center}
    \begin{tabular}{l}
        a  \\
        \includegraphics[width=2.25in]{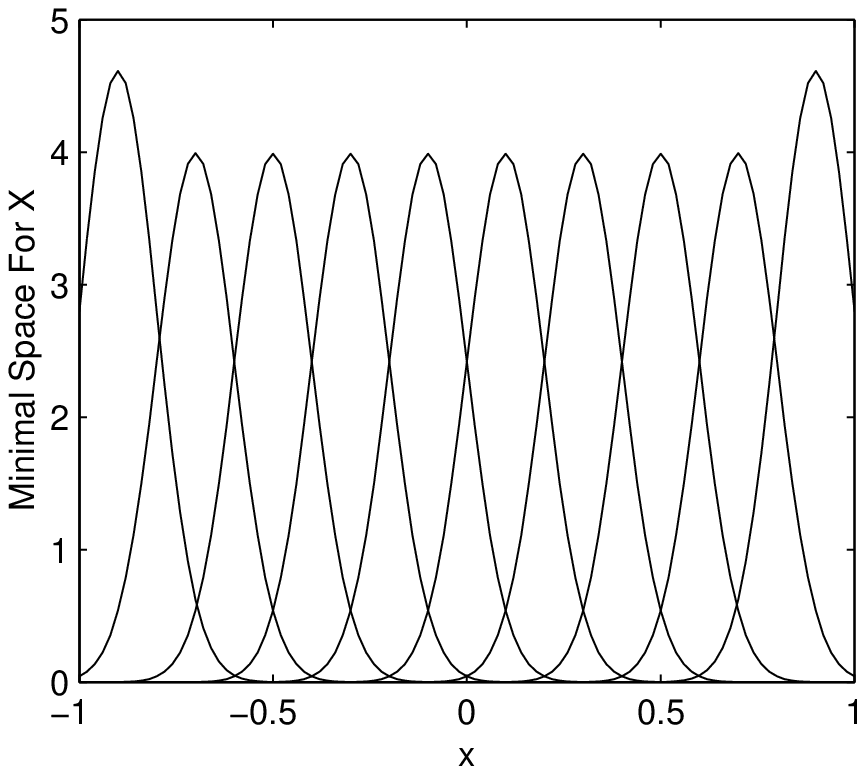} 
    \end{tabular}
    \hfill
    \begin{tabular}{l}
	b  \\
	\includegraphics[width=2.25in]{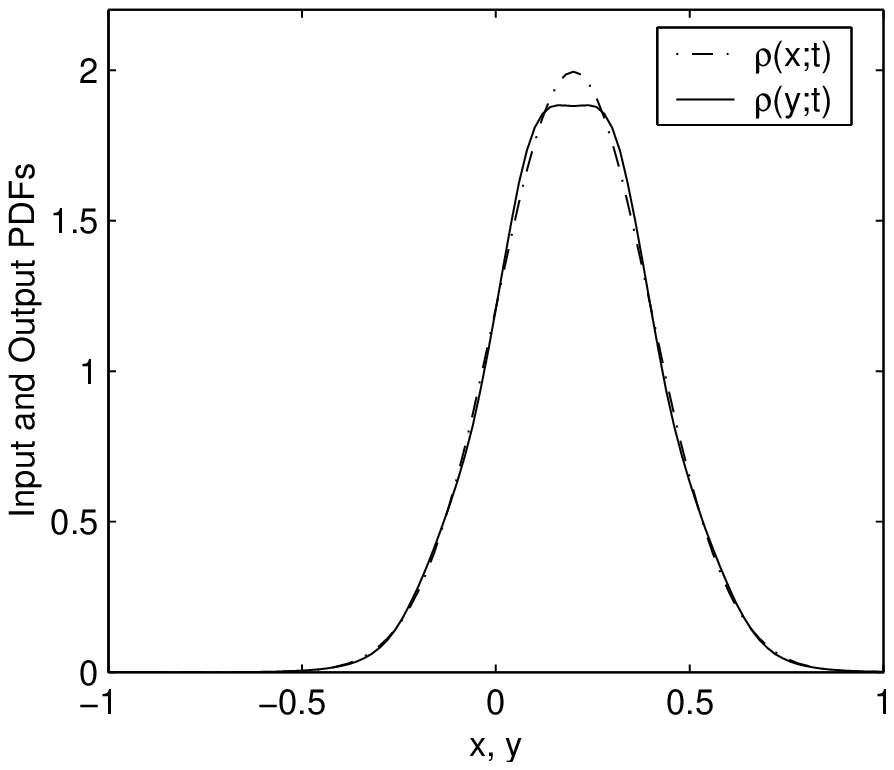} 
    \end{tabular}
    \begin{tabular}{l}
	c  \\
	\includegraphics[width=2.25in]{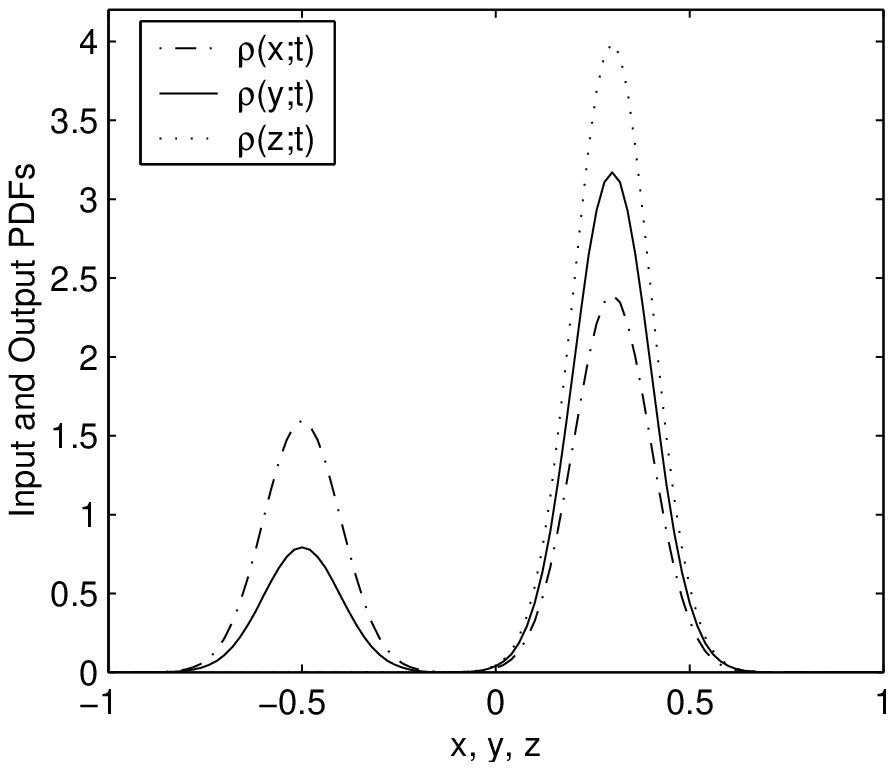} 
    \end{tabular}
    \hfill
    \begin{tabular}{l}
	d  \\
	\includegraphics[width=2.25in]{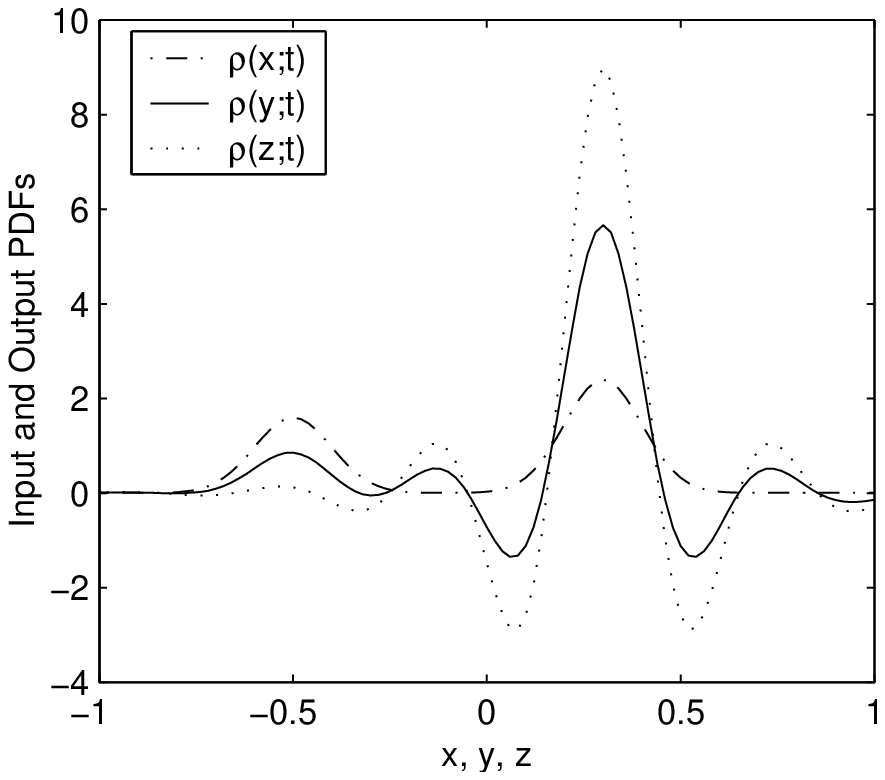} 
    \end{tabular}
    \end{center}
    \caption{Propagation of evidence in chain-structured neural belief 
    networks.  (a) Gaussian functions spanning the minimal spaces for the
    chain-structured neural belief networks under consideration.  
    Orthonormal bases for the minimal spaces are formed from 
    these functions.
    (b) A neural belief network accurately transmits PDFs in a chain 
    of two nodes.  The decoded output PDF $\rho(y;t)$ is very similar to the 
    encoded input PDF $\rho(x;t)$.
    (c) Multiple sources of evidence can help to resolve ambiguous 
    information.  Here, the inferior mode of a bimodal, bottom-up input 
    $\rho(x;t)$ is damped by a more specific top-down signal 
    $\rho(z;t)$.
    (d) A high-level model can be dynamically generated in a neural 
    belief network with a population $Z$ of winner-take-all neurons.  An 
    ambiguous input signal at $X$ is propagated to the winner-take-all 
    neurons through $Y$.  The winner-take-all neurons only respond to the 
    stronger mode of the bimodal input, which damps the inferior mode in 
    $\rho(y;t)$.  
    The functions \( \rho(y;t) \) and~\( \rho(z;t) \) are 
    only approximations to PDFs, since they can take on negative values.
    }
    \label{fig:threenodechain}
\end{figure}

For the chosen demonstration, we employ representations of higher 
dimensionality than those generally used in the preceding 
developments.  For each graph node, the interval $[-1, 1]$ is covered
with ten Gaussians (Figure~\ref{fig:threenodechain}a), from which
an orthonormal basis is generated.  This makes it possible to
represent a variety of PDFs, including multimodal distributions.

The neural firing rate
patterns in the explicit space are also assumed to be of Gaussian form.  To 
eliminate any possibility of negative firing rates, the 
nonlinear activation function is taken to be rectification.

We have already studied at some depth the behavior of chains consisting 
of two nodes (sections~\ref{sec:inferencenbns} 
and~\ref{sec:retrosupport}).  These chains 
 are able 
to transmit probabilistic information from one node of the graph to 
another, with the accuracy limited by the representations adopted.  For 
the predictive network, a feedforward input $X$ drives the 
output $Y$.

To keep the focus on the interaction of multiple sources of evidence, we take 
the conditional probability to be a Dirac delta function, 
$\rho(y\given x) = \delta(y - x)$.  Of course, this defines a version of 
the communication channel that we have extensively considered, but the 
goal is now to duplicate the PDF rather than a particular numerical 
value.  As seen in Figure~\ref{fig:threenodechain}b, the input PDF $\rho(x;t)$  is 
copied with great fidelity by the output PDF $\rho(y;t)$.

Every chain with three or more nodes will admit both predictive and 
retrospective support.  The behavior of all such chains is well 
characterized by a chain with three nodes: the update rule for any 
node depends only on the neighboring nodes, so a chain with three 
nodes covers all possibilities (only predictive support, only 
retrospective support, and both predictive and retrospective support).  
Therefore, we consider the effect of adding a third node to the chain 
(Figure~\ref{fig:trees}d).
We set $\rho(z\given y) = \delta(z - y)$ along with $\rho(y\given x) = 
\delta(y - x)$ and use identical  parameters $\relstr_{y}$ and 
$\relstr_{z}$.

The first possibility is just to add the third node
without adding any additional evidence to the network.  We decode 
$\rho(y;t)$ and $\rho(z;t)$ from neural firing rates determined using 
update rules derived previously for more general trees.  For the same input 
PDF as above, $\rho(y;t)$ is  
unchanged by the addition of the retrospective support from the third 
node, and the structure of $\rho(z;t)$ is identical to that of $\rho(y;t)$. 

Since the same evidence was presented to the network, 
the introduction of a third node did not change the behavior of the 
original nodes.  This is entirely appropriate, given the probabilistic 
foundations of the neural belief networks.  However, it is feasible 
that, by adjusting the \( \relstr_{i} \) parameters or the PDFs 
representable by the minimal spaces, there can be a change in the dynamics 
of neural networks extended in this fashion.  For example, appropriate 
parameter choice  could produce a slowly varying PDF in $Z$ that is 
relatively stable against noise, thus stabilizing a more rapidly varying 
PDF in~$Y$. 

A more proactive role that the third node $Z$ can play is as an 
additional source of evidence.  We directly specify $\rho(z;t)$ and 
encode it into the neural network.  The neural firing rates for $Y$ are 
driven by predictive support from $X$ and retrospective support from 
$Z$, and then decoded to find $\rho(y;t)$.  One possible use of this 
second source of evidence is to resolve an ambiguous input; the inferior 
mode of a bimodal predictive input can be de-emphasized using more 
specific retrospective evidence (Figure~\ref{fig:threenodechain}c).

Although this use of the retrospective evidence resolves the ambiguous 
predictive input, it does not 
explain how the retrospective evidence comes about.  Ideally, we would 
like to build up a high-level model of the predictive input, and use 
this model to generate a top-down signal that imposes global 
regularity on the bottom-up predictive input.  The uniform priors that 
we have assumed throughout this work are somewhat of a 
hindrance for this purpose.  With these priors, probabilities account well for multiple 
possibilities, but here we want to choose only one of those 
possibilities.

In principle, we could restrict the allowed PDFs  for $\rho(z;t)$ 
by imposing a prior on the PDFs and rederiving our neural update rules.  
However, we will adopt a more direct approach that illustrates the 
first steps towards implementing decision theory \cite{raiffa:1968}
in neural networks 
and take the neurons 
representing $\rho(z;t)$ to be winner-take-all units.  Only one of 
these neurons will be active at a time, based on a simple utility 
function:  the neurons compete to be 
active, and the single neuron that is most strongly driven will be the 
one that is activated.  The manner of implementation of the winner-take-all 
units is immaterial, so we directly choose the most strongly driven neuron in 
the computer simulations.  (It is possible to implement a set of 
winner-take-all units in a real network through a suitable choice of 
nonlinear activation function and lateral weights.  One may let each 
neuron inhibit the others and have a self-excitatory connection.  
See Hertz \etal, 1991\nocite{hertz/etal:1991}.)

Since only the most strongly driven winner-take-all neuron is 
activated, this strategy provides us with a way to generate a 
high-level model that selects the dominant mode of a multimodal input 
distribution.  We allow $Z$ to be driven by the predictive support 
from $Y$, and the winner-take-all nature of the $Z$ neurons permits 
only narrow Gaussian PDFs to be represented.  Thus, $\rho(z;t)$ serves 
as our high-level model, and can resolve ambiguous inputs, as demonstrated in 
Figure~\ref{fig:threenodechain}d. 
This type of neural network could be used as the starting point for  
coherent theoretical accounts of 
attentive effects in the primate visual system, of the electrosensory system of 
weakly electric fish, and of other neural systems where an internal model 
is built up to impose global constraints on neural representations of 
information.

\section{Conclusions}

We have
extended
the hypothesis that neural 
networks represent information as probability density functions.  
These PDFs are assumed to be obtainable by a linear combination of some 
implicit decoding functions, with the decoder for each neuron being weighted 
by its firing rate.  The firing rates in turn are obtainable from 
the PDF using a complementary set of encoding functions.

Success in representing an individual PDF with a population of 
neurons \cite{barber/clark/anderson:2001a} has led us 
 to inquire whether more complex probabilistic 
models can be represented and implemented in neural terms.  To this 
end, we have adopted graphical representations of 
probabilistic dependence models, in the form of Pearl's \bbns, 
to organize and simplify the relations between the random variables 
in such a model.

In summary, we have introduced, analyzed, and applied three ways to 
represent probabilistic information.  These are
the implicit model, depicted as a \bbn; the representation 
of the probabilistic model in the minimal space; and the 
representation of the probabilistic model as a neural network in the 
explicit space.  Further, we have devised rules that permit us to convert 
one type of representation into another type.  This formalization
of the representation and processing of information in 
neurobiological computation therefore suggests a general protocol 
by which probabilistic models can be 
embedded in neural networks.  First, we specify the probabilistic 
model, using a \bbn\ to organize the random variables.  We then 
consider what sorts of functions are exemplars of the PDFs describing 
the random variables 
and use these 
exemplars to define the minimal space.  Finally, we utilize the 
relations between the minimal space and the explicit space to 
generate the neural network itself.

\section*{Acknowledgements}

This research was supported by the National Science Foundation under
Grant Numbers IBN-9634314, PHY-9602127, and PHY-9900713.  
While writing this work, MJB was a member of the Graduiertenkolleg 
Azentrische Kristalle, GK549 of the DFG; the writing was completed 
while MJB and JWC were participants 
in the Research Year on ``The Sciences of Complexity: From Mathematics 
to Technology to a Sustainable World'' at the Zentrum f\"ur 
interdisziplin\"are Forschung, University of Bielefeld.

\bibliographystyle{apalike}
\bibliography{neurcomp,math,mjbpubs}

\end{document}